\documentclass{article}

\usepackage{amsfonts}
\usepackage{amssymb}
\usepackage{amsmath}
\usepackage{amsthm}
\usepackage{mathtools}
\usepackage{graphicx}
\usepackage{amscd}
\usepackage{mathrsfs}
\usepackage{tipa}

\begin{document}

\title{Degrees of Freedom in modified Teleparallel Gravity}

\author{Alexey Golovnev\\
{\small \it Centre for Theoretical Physics, The British University in Egypt,}\\
{\small \it El Sherouk City, Cairo 11837, Egypt}\\
{\small  agolovnev@yandex.ru}} 
\date{}

\maketitle

\begin{abstract}

I discuss the issue of degrees of freedom in modified teleparallel gravity. These theories do have an extra structure on top of the usual (pseudo)Riemannian manifold, that of a flat parallel transport. This structure is absolutely abstract and unpredictable (pure gauge) in GR-equivalent models, however it becomes physical upon modifications. The problem is that, in the most popular models, this local symmetry is broken but not stably so, hence the infamous strong coupling issues. The Hamiltonian analyses become complicated and with contradictory results. A funny point is that what we see in available linear perturbation treatments of $f(T)$ gravity is much closer to the analysis with less dynamical degrees of freedom which has got a well-known mistake in it, while the more accurate work predicts much more of dynamics than what has ever been seen up to now. I discuss possible reasons behind this puzzle, and also argue in favour of studying the most general New GR models which are commonly ignored due to suspicion of ghosts.

Prepared for the Proceedings of XII Bolyai–Gauss–Lobachevsky Conference (BGL-2024): Non-Euclidean Geometry in Modern Physics and Mathematics.

\end{abstract}

\section{Introduction}

These days, modified gravity is very popular due to a variety of reasons ranging from purely phenomenological troubles to deep theoretical issues. None of these motivations to modify it are unquestionable. However, the current situation is puzzling enough for a big flow of modified gravity papers to go on. An interesting point about this business is that it turns out very difficult to non-trivially modify general relativity (GR) without creating a catastrophe, of one sort or another, which infortunately very often goes unaddressed for the sake of producing more papers.

Among the interesting options on the market, there is an old idea of radically chaging geometry of spacetime through adding, on top of the metric, yet another connection which is flat. These teleparallel approaches can safely start from simply reproducing GR in a very unnatural language. Specifying the notion of a flat connection to two extreme cases of torsion only and non-metricity only, one can talk about the trinity of gravity \cite{trinity}. Since, in the real world, all the test particle trajectories do correspond to the Riemannian geodesics of the observable metric, I would not go for this point of view. In any other theory, one can also introduce some new hidden structures and use them for constructing another version of the known physics.

Going even much farther in employing unobservable artificial structures, one can claim having solved the problem of energy in gravity \cite{ener1, ener2}. I strongly disagree with that \cite{pamphlet}, unless we introduce a substantially modified teleparallel model instead of a GR-equivalent one. In case of simply reproducing GR, the flat connection is completely esoteric for a mundane observer, and one could indeed do many different constructions for the same goal, such as introducing a fixed Minkowski space and treating the real-world metric as a dynamical field on top of that. It is a clear way of having well-defined conservations laws, though in relation to artificial pillars built by nothing but our imagination. I prefer to admit that generically there are no objective notions of conserved energy and the like.

I would also say that the teleparallel descriptions of gravity have no clear relation to the real world. In GR-equivalent models, just every flat connection goes, modulo constraints which might be imposed by the very definition of the model at hand, e.g. vanishing non-metricity for metric teleparallel or vanishing torsion for symmetric teleparallel. However, they provide us with novel ways of modifying the theory of gravitational interactions. This is interesting in its own right, let alone giving us ways of better understanding GR and its place in the theory landscape.

In Section 2, I briefly describe the general concept of teleparallel geometry, then the zero non-metricity condition will be imposed for the rest of the paper. In Sections 3 and 4, I discuss the most general (though parity-preserving) New General Relativity (New GR) models. Usually such options are disregarded due to the fear of ghosts \cite{old}. My claim \cite{us} is that this question should be investigated in much better detail. If anything in metric teleparallel gravity, it is this case which can meaningfully help us with conservation laws, for the full structure of the flat parallel transport becomes physical. Then, in Sections 5 and 6, I discuss probably the most popular modified teleparallel gravity, the $f(\mathbb T)$ one \cite{fT}. This is a simple modification of the GR-equivalent case, and the choice of the teleparallel geometry is no longer free in it, but it is not fully fixed either, leading to the intricate zoo of remnant symmetries \cite{remn}, ubiquitous strong coupling issues \cite{FoundIss}, and therefore an ill-defined number of degrees of freedom. Despite being a total theoretical disaster, it is still very actively used for (naive) phenomenology. Finally, in Section 7, I conclude.

\section{Teleparallel geometry}

Let me start from the basic notion of teleparallel structures. Namely, we assume that, on the spacetime manifold, there exists an independent connection of vanishing curvature tensor. Since the curvature tensor describes a change in a vector field upon parallelly transporting it over an infinitesimal closed contour, it means that if one transports a vector from one point to another over two different smooth trajectories, the result is the same as long as those paths are smoothly deformable into each other. Modulo global obstructions in case of nontrivial topology, we therefore get an unambiguous notion of two vectors at a distance being equal, or parallel to each other, and hence the name.

All in all, given a flat parallel transport, one can choose a basis of 1-forms $e^a=e^a_{\mu} dx^{\mu}$ at some point and get a covariantly constant basis of 1-forms $e^a_{\mu} (x)$ over the full spacetime, or at least over a topologically trivial patch around the initial point. With the global freedom of initially choosing the basis, this "proper" co-tetrad, or a set of covariantly constant 1-form fields, is a faithful representation of the teleparallel connection. At the same time, one can also go for the dual basis of $\mbox{\textschwa}_a^{\mu}$ of covariantly constant vectors, or a tetrad, with $\mbox{\textschwa}=e^{-1}$ in terms of matrices. 

Note that usually I denote both tetrads and co-tetrads with the same letter $e$ leaving the distinction between them solely for the position of the Latin and Greek indices. For the pedagogical purposes, here I follow the notation of letters $\mbox{\textschwa}$ and $e$ from the classical trinity paper \cite{trinity}. Another option available in the literature \cite{MG1} is to use the letters $E$ and $e$.

All in all, a teleparallel geometry does have a basis of covariantly constant vector fields, or a soldering form of vanishing spin connection in the "covariant" language, $\partial_\mu e^a_{\nu} - \Gamma^{\alpha}_{\mu\nu}e^a_{\alpha}=0$ which implies an affine connection of the Weitzenb{\"o}ck type
\begin{equation}
\label{Weitzen}
\Gamma^{\alpha}_{\mu\nu}=\mbox{\textschwa}_a^{\alpha}\partial_{\mu} e^a_{\nu}.
\end{equation} 
If we believe that the flat connection (\ref{Weitzen}) does objectively exist on the given spacetime manifold, then its defining (co-)tetrad $e^a_{\mu}$ is not free to be chosen. To the contrary, it is a dynamical variable and must be ruled by equations of motion of a model at hand. In case we want to deal with an arbitrary (orthonormal) tetrad, for either describing an observer or coupling the fermions, it must be another object, say $h^a_{\mu}$, which serves then as yet another basis for representing all the geometric quantities.

All the consideration above does not depend on a particular type of teleparallel models. One can specify it further, and it brings us to the two basic curvatureless frameworks of the trinity \cite{trinity}. The simplest version, even though it appeared much later than the classical teleparallel, is called {\bf symmetric teleparallel}, and this is about a (tele)parallel transport with no torsion either, $\partial_{\mu} e^a_{\nu}=\partial_{\nu} e^a_{\mu}$. At least locally, the basic co-tetrad can then be represented as a coordinate basis,
\begin{equation}
\label{symtet}
e^a_{\mu}=\frac{\partial\zeta^a}{\partial x^{\mu}}.
\end{equation}
In other words, the structure of the parallel transport is that of a Minkowski space, with $\zeta^a$ scalar fields being its Cartesian coordinates. It's just that the physical metric is a different field, and therefore non-trivial non-metricity $Q_{\alpha\mu\nu}\equiv\bigtriangledown_{\alpha}g_{\mu\nu}$ is there.

To put it yet another way, there are $\zeta^a$ coordinates (\ref{symtet}) with vanishing affine connection coefficients in the symmetric teleparallel framework. The GR-equivalent model in this case (STEGR) is basically given by the non-covariant $\Gamma\Gamma$ action of Einstein with the partial derivatives of the metric interpreted as components of the non-metricity tensor in the Cartesian coordinates of the teleparallel structure.

Another option is the classical or {\bf metric teleparallel} geometry in which we allow for torsion only. This is the framework of this paper. The condition of vanishing non-metricity means then that all the scalar products do not change upon the parallel transport. In particular, we can consistently choose the defining tetrad to be orthonormal. The usual approach to that is to treat the tetrad as the only dynamical variable from which the metric is \underline{defined} by
\begin{equation}
\label{metr}
g_{\mu\nu}=\eta_{ab} e^a_{\mu} e^b_{\nu}.
\end{equation}

In this case, two different letters for tetrads and co-tetrads look particularly funny, for going from one to another can be considered as simply raising and lowering the Greek indices by the spacetime metric and Latin ones by the Minkowski one. We then probably have to also write the inverse metric as $\rotatebox{180}{g}^{\mu\nu}=\eta^{ab} \mbox{\textschwa}_a^{\mu} \mbox{\textschwa}_b^{\nu}$.

The teleparallel connection (\ref{Weitzen}) is automatically compatible with the metric (\ref{metr}) while it does generically have a non-trivial torsion tensor
\begin{equation}
\label{tor}
{T^{\alpha}}_{\mu\nu} = \Gamma^{\alpha}_{\mu\nu} - \Gamma^{\alpha}_{\nu\mu},
\end{equation}
and one can easily check that the torsion scalar
\begin{equation}
\label{tegr}
{\mathbb T}=\frac14 T_{\alpha\mu\nu}T^{\alpha\mu\nu} + \frac12 T_{\alpha\mu\nu}T^{\mu\alpha\nu} - T_{\mu}T^{\mu}\qquad \mathrm{where} \qquad T_{\mu}\equiv {T^{\alpha}}_{\mu\alpha}
\end{equation}
differs from (minus) the usual (Levi-Civitian) scalar curvature by only a surface term. Therefore, the Lagrangian density of $\mathbb T$ defines a teleparallel theory equivalent to general relativity (TEGR).

\section{New General Relativity}

Having fixed the metric teleparallel framework of formulae (\ref{Weitzen}) and (\ref{metr}), i.e. a flat and metric-compatible connection, one of the most natural ideas \cite{HaSh} in the quest for modified gravity is to modify the coefficients in the torsion scalar (\ref{tegr}):
\begin{equation}
\label{ngr}
{\mathfrak T}= \frac{a}{4}\cdot T_{\alpha\mu\nu}T^{\alpha\mu\nu} + \frac{b}{2}\cdot T_{\alpha\mu\nu}T^{\mu\alpha\nu} - c\cdot T_{\mu}T^{\mu}.
\end{equation}
This is the most general quadratic in torsion (and parity-preserving) invariant. The action of $\int {\mathfrak T}\cdot  \sqrt{-g} d^4 x$ defines what is known as New GR, with restoration of the good old GR in case of $a=b=c$.

Historically, the case of the so-called one-parameter New GR \cite{HaSh}, that is $a+b=2c$ with one more free parameter removed by fixing the effective gravitational constant, is very much preferred over other cases due to the claimed absence of ghosts \cite{old} and the same static spherically symmetric solutions as in GR \cite{HaSh}. However, on one hand, this option seems to lack much interest, for deviations from GR can hardly be seen, neither for (unperturbed) astrophysical solutions \cite{sphrwe} nor in linear cosmological perturbations \cite{cosmwe}, unless one goes for unnaturally complicated tetrads in the background solutions \cite{cumbersome}. On the other hand, there are arguments that the dynamical structure of such restricted models cannot be robust and stable \cite{obstr}.

In this respect, my opinion \cite{us} is that the most general ("type 1") New GR models (\ref{ngr}), that is
\begin{equation}
\label{gennew}
a \neq b, \qquad a \neq -b, \qquad a+b \neq 2c, \qquad a+b \neq 6c,
\end{equation}
are the most promising ones. Practically, there are no remnant symmetries. Out of sixteen variables, four are pure gauge due to diffeomorphism invariance, four more are physical but constrained, due to the gauge symmetries "hitting twice", and therefore eight dynamical modes are present. In both Minkowski \cite{us} and (spatially flat) cosmological spacetimes \cite{cosmwe}, all the polarisations of waves are clearly seen.

Let me briefly summarise the simplest, weak gravity case. In order to study gravitational waves around the trivial Minkowski background, $e^a_{\mu}=\delta^a_{\mu}$, one needs to consider the most general perturbation of the tetrad, as opposed to only some possible choice of a tetrad for the most general perturbation of the metric. Modified teleparallel gravities have got more equations of motion, due to their non-trivial antisymmetric part, and they require more variables of course. Or in other words, the local Lorentz invariance is broken, and therefore various tetrads for the same metric are physically different. In particular, in $f({\mathbb T})$ gravity the Lorentz boosts of the perturbed tetrad do influence the scalar cosmological perturbations \cite{GKcosm}. If one goes for Lorentz covariant description, then the new variables in the spin connection must be taken into account \cite{mespin}.

I continue in the pure-tetrad approach \cite{us}. Separating the perturbations into scalars, divergenceless vectors, and a symmetric traceless divergenceless tensor, one can parametrise the perturbed tetrad as
\begin{equation}
\label{pert}
\begin{array}{rcl}
e^0_0 & = & 1 + \phi,\\
e^0_i & = & \partial_i \beta +{\mathcal L}_i +{\mathcal M}_i,\\
e^i_0 & = & \partial_i \zeta+{\mathcal L}_i -{\mathcal M}_i,\\
e^i_j & = & (1 -\psi) \delta_{ij}+\partial^2_{ij}\sigma+\epsilon_{ijk}\partial_k s+\partial_j c_i+\partial_i\chi_j - \partial_j\chi_i+\frac12 h_{ij}.
\end{array}
\end{equation}
where the perturbations of the metric tensor are given by $\phi$, $\psi$, $\beta-\zeta$, $\sigma$, $2{\mathcal M}_i$, $c_i$, and $h_{ij}$. On top of that, we have got (\ref{pert}) Lorentz boosts in $\beta+\zeta$ and $2{\mathcal L}_i$, and spatial rotations of $s$ and $\chi_i$. Given that diffeomorphism invariance is still there (with the tetrad taken as a set of vectors), we should fix a gauge which will be \cite{us, cosmwe, GKcosm}
\begin{equation}
\label{gauge}
\beta=\zeta, \qquad \sigma=0, \qquad c_i=0
\end{equation}
leaving us with twelve physical modes as long as the condition (\ref{gennew}) is satisfied.

There is no constraint in the tensor sector. If $a\neq -b$, there is no new gauge freedom either; and the two standard polarisations of a graviton do obey the simple wave equation
\begin{equation}
\label{tensor}
{\ddot h}_{ij} -\bigtriangleup h_{ij}=0
\end{equation}
which goes away if $a+b=0$.

If $a\pm b \neq 0$ and $a+b \neq 2c$, the vector sector can be represented as
\begin{equation}
\label{vector}
{\ddot{\mathcal M}}_i- \bigtriangleup{\mathcal M}_i =0, \qquad {\ddot\chi}_i - \bigtriangleup \chi_i = \frac{2b(a+b)-4ac}{(a-b)(a+b-2c)}\cdot {\dot{\mathcal M}}_i, \qquad {\mathcal L}_i=\frac{a+b}{a-b}\cdot {\mathcal M}_i - {\dot\chi}_i
\end{equation}
where the first equation initially had a factor of $a+b$. In the three divergenceless vectors, there are six variables. We see that two modes are constrained (the last equation), while four of them are dynamical, with only two dynamical modes being visible in the metric (${\mathcal M}_i$).

Finally, in the scalar sector, the new variables $s$ and $\zeta$ are not pure gauge if $a\neq b$ and $a+b\neq 2c$ respectively. If also $a\neq -b$, the scalar mode in the metric must be conformal, $\phi=-\psi$, and finally we get
\begin{equation}
\label{scalar}
{\ddot s} - \bigtriangleup s =0, \qquad {\ddot\zeta} - \bigtriangleup\zeta =0, \qquad \phi=-\psi=\frac{2c-a-b}{6c-a-b}\cdot \dot\zeta.
\end{equation}
There are two dynamical modes. One of them ($s$) is hidden from the usual observers, while another one ($\zeta$) presents itself in the conformal mode of the metric.

All in all, there are eight dynamical modes, five of which are visible in the metric (\ref{tensor}, \ref{vector}, \ref{scalar}). What can be seen in the metric is somewhat similar to the ghost-free massive gravity: one tensor, one vector, and one scalar. On top of that, there are three dynamical modes residing purely in the Lorentz group. The four constraints do fix another half of the Lorentz, and also impose one restriction ($\phi=-\psi$) on the six metric variables.

One can also look at the New GR framework as a theory which has four vector fields (composing the tetrad) and with the action which is quadratic in derivatives coming only in combinations (\ref{tor}) of 
\begin{equation}
\label{comb}
{\mathfrak F}^a_{\mu\nu}=\partial_{\mu}e^a_{\nu}-\partial_{\nu}e^a_{\mu}.
\end{equation}
As long as the corresponding gauge symmetries are preserved, we normally expect eight degrees of freedom, unless there are some extra fine tunings of parameters. This is precisely what happens in the quadratic weak gravity action above. Non-linearly, the $U(1)^{\bigotimes 4}$ symmetry is no longer there, and one would generically expect then twelve degrees of freedom. However, the Abelian symmetry gets replaced by full diffeomorphisms which still reduce the number of dynamical modes by four due to a generalisation of Bianchi identities \cite{HaSh, sphrwe, Bianchi}.

\section{Instabilities of generic New GR?}

Coming back to stability issues, a paper \cite{them} appeared recently analysing the New GR models more carefully than it used to be with the classical result \cite{old} which had been obtained by the use of spin projection operators (hence new derivative terms) directly inside the action. As is mentioned above, this old work \cite{old} led the community to accepting only a 1-parameter model of New GR, or more precisely a 2-parameter one if not to insist on the measured value of effective gravitational constant. Of course, extra derivatives in the action generically do change a model at hand. Recently, we claimed with colleagues \cite{us} that the generic 3-parameter New GR theory might actually be healthy, while the new paper \cite{them} asserts that it's not, due to ghosts in the vector sector of perturbations.

To be honest, I cannot tell it for sure, for the full answer would require thorough analysis of all the dynamical features, ideally beyond the linear approximation. What can be seen from the paper \cite{us} is that the kinetic part of the (gauge-fixed linearised) Hamiltonian can easily be positive definite, except for the non-dynamical conformal mode. For sure, it is not enough for really implying stability. For example, a massive vector field with $m^2<0$ does have a positive definite kinetic part of its canonical Hamiltonian, too. However, it ceases to be the case upon solving for the temporal component or using the St{\"u}ckelberg trick. Nevertheless, it must always be accurately analysed, while the argument of the paper \cite{them} does not justify the claim.

A minor issue is that they are going for an action in terms of gauge-invariant variables. As I explained elsewhere \cite{anti}, the gauge structure is crucial. One might indeed go for Lorentz-gauge-invariant variables in covariant teleparallel theories thus producing the pure-tetrad ones \cite{weL}, with no change to the physical content. However, this is so due to the purely algebraic nature of the symmetry, which is not the case of diffeomorphisms. Turning the blind eye to fundamental aspects, changes in the numbers of spatial derivatives are often quite benign for the perturbation theory where we, say, put any harmonic function to identically vanish, however it can produce essential differences when involving time derivatives. For example, in electrodynamics, taking the gauge-invariant field strength as the dynamical variable removes derivatives from the action making the equations trivial.

An important point in understanding the paper \cite{them} is that they think only in terms of dynamical ("propagating") modes, as if the constrained ones were not physical. In electrodynamics, it's possible to go for an action for the transverse (gauge-invariant) modes only. It can be done indeed, but it is not the same as real electrodynamics in which the longitudinal mode is also physical, for it has the Coulomb's law in it. In case of New GR, this attitude also makes them think about potential viability of gravity models \cite{them} in terms of propagating ghosts only, without caring of whether an accidental gauge freedom (beyond diffeomorphisms) appears in the metric sector. It is not a good idea, of course. It is all right if some of the metric perturbations are not propagating by themselves, but they must be predictable, one way or another, in order for the usual coupling of matter to make sense. An extra gauge freedom in metric perturbations is not admissible.

All in all, I must admit that the count of dynamical modes \cite{them} in the type 1 vector sector went correct. In their notations, the only time derivative in this part of the trick was in defining a new gauge-invariant variable $D_i = S_i - {\dot F}_i$. The variation with respect to $D$ is then equivalent to the variation in terms of $S$. Analogously, we can imagine a mechanical system for $x(t)$ and $y(t)$ with a Lagrangian $L=(y-\dot x)^2$ with the equation of $y=\dot x$. One can take the gauge invariant variable $Y=y-\dot x$ for which the Lagrangian $L=Y^2$ demands $Y=0$. This equation is the same as it was before, from the variation with respect to $y$. At the same time, it is more restrictive than it was in the equation for $x$, however the latter was anyway overtaken by the equation for $y$.

In the end of the day, they have found \cite{them} two fully dynamical transverse vectors, and one transverse vector constrained. This is the same result (\ref{vector}) as we had \cite{us}. Then the ghostly claim \cite{them} comes from the fact that the Authors were able to derive a fouth-order equation from the two second-order ones. It is absolutely incorrect to deduce the ghosts from that. It's not a big deal, to do so for many absolutely stable systems.

Let me discuss a simple toy model of the Lagrangian
$$L=\frac12 \left({\dot x}^2 + {\dot y}^2 - 2xy\right)$$
which corresponds to a Hamiltonian
$$H=\frac12 \left(p_x^2 + p_y^2 + 2xy\right)$$
with a positive definite kinetic part and no constraints in it. There is no ghost at all, even though the potential energy is not bounded. Nonetheless, the equations of motion
$$\ddot x + y =0, \qquad \ddot y + x =0$$
can immediately be brought to the form of
$$\ddddot x - x =0, \qquad y=-\ddot x$$
which is a fouth-order equation for one of the variables, with the second one being uniquely determined then. It does not mean that we have produced a ghost out of nowhere.

In case of worries, one can go for $\frac12 (x+y)^2$ instead of $xy$ in the potential energy which still allows for the same higher-derivative-order rewriting, even though it would look far less natural than a simple change of variables to $x+y$ and $x-y$. Or, to make it even sharper, one can start from an absolutely stable model of $L=\frac12 \left({\dot x}^2 + {\dot y}^2 - x^2\right)$ and introduce new variables $s=\frac{x+y}{\sqrt{2}}$ and $a=\frac{x-y}{\sqrt{2}}$. It is then easy to rewrite the equations as a constraint for one of them and a fourth-order equation for the other. And indeed, two modes of second-order equations need four Cauchy data altogether, therefore the whole freedom can be represented in terms of a single fourth-order equation. Does it really mean a bad ghost in the system?

Therefore, the claim \cite{them} of having found ghosts is unsubstantiated. Of course, at the linear level, even the question of stability is not very meaningful, for the dynamics is fully under control. Once we turn interactions on, in a Lorentz-invariant theory with ghosts, negative kinetic energies generically yield an infinite volume of ways to produce new excitations respecting all conservation laws, hence we usually expect to see an instability with no finite time scale. This is a very interesting question, whether the ghosts are present or not, even if we can still live with some of them in a physical theory \cite{Vikman}, but it requires a more detailed analysis. What I am quite sure about is that the generic (type 1) models \cite{us} are pretty robust in terms of their physical modes. It's always four diffeomorphisms hitting twice and no more constraints. The issue of stability is yet to be investigated.

\section{Non-linear $f(\mathbb T)$ models}

Let me now turn to a very popular model, $f(\mathbb T)$ gravity. We come back to the standard torsion scalar (\ref{tegr}) of TEGR, and put a non-liear function of it into the action: $\int f({\mathbb T})\cdot  \sqrt{-g} d^4 x$. The torsion scalar $\mathbb T$ itself was almost the usual Einstein-Hilbert Lagrangian density, different from the latter by only a surface term which does not change anything in the system of equations. However, once a surface term has got into the argument of a non-linear function, it ceases to be such. Still, one can use this structure of the action for facilitating the variations and subsequent calculations a lot \cite{myissues}.

The equations of motion (in vacuum) are worth to be written in a covariant form
\begin{equation}
\label{fTeom}
f_{T}\cdot G_{\mu\nu}+\frac12 \left(f-f_{T}{\mathbb T}\right)\cdot g_{\mu\nu}+f_{TT}\cdot S_{\mu\nu\alpha}\partial^{\alpha}{\mathbb T}=0,
\end{equation}
as opposed to big parts of modified teleparallel literature. A few comments on the notations are in order. The superpotential, or a torsion conjugate, $S_{\alpha\mu\nu}$ is a tensor which can be defined
\begin{equation}
\label{spot}
S_{\alpha\mu\nu}=\frac12\left(K_{\mu\alpha\nu}+g_{\alpha\mu}T_{\nu}-g_{\alpha\nu}T_{\mu} \right), \qquad K_{\alpha\mu\nu} =\frac12 \left(T_{\alpha\mu\nu}+T_{\nu\alpha\mu}+T_{\mu\alpha\nu}\right)
\end{equation}
in terms of the contortion tensor $K_{\alpha\mu\nu}$ (\ref{spot}) which, in turn, is a difference between the teleparallel connection and the Levi-Civita one. The $G_{\mu\nu}$ is the usual Einstein tensor calculated from the metric $g_{\mu\nu}$. Since it goes then in terms of the Levi-Civita connection, we often denote it by $\mathop{G_{\mu\nu}}\limits^{(0)}$ or $\mathring{G}_{\mu\nu}$. Finally, there are derivatives of the function $f$, i.e. $f_T\equiv\frac{df}{d\mathbb T}$ and $f_{TT}\equiv\frac{d^2 f}{d\mathbb T^2}$.

We immediately see that, in case of a linear function $f(\mathbb T)$, the equations (\ref{fTeom}) reduce to those of GR, or TEGR, with only the gravitational constant renormalised by the factor of $f_T$ and an additional cosmological constant given by $f(0)$. Genuinely new features are solely brought about by the $f_{TT}$ term (\ref{fTeom}). There is no surprise in that, since what is responsible for non-trivial modifications of gravity is precisely the non-linearity of the function $f$. Note also that this is the only term in the equations (\ref{fTeom}) which has got a non-trivial antisymmetric part and also depends on Lorentz rotations of the tetrad beyond the scalar coefficients. Of course, the reason for that is the broken local Lorentz invariance. A bit more worrisome is that the constant $\mathbb T$ solutions do never go beyond the limit of GR (except the pathological $f_T =0$ cases), and we should probably be able to find a (very often quite unnatural) Lorentz transformation of a simple tetrad which would do it for any given metric, as for instance was done (using null tetrads) for Kerr \cite{T01} and cosmological \cite{T02} metrics.

So far so good, but then the foundational issues come \cite{FoundIss}. As has already been mentioned above, one and the same metric can correspond to different tetrads, even if orthonormal ones. When those are physical objects in themselves, we face the issue of choosing the real physical one. In case of weak gravity, it is natural to take the trivial tetrad of $e^a_{\mu}=\delta^a_{\mu}$. It is the one which respects all the symmetries and also enjoys vanishing torsion tensor. What else would we want from the true vacuum? However, then the torsion scalar (\ref{tegr}) is already quadratic in perturbations around this important background. Hence, the quadratic action feels only the linear term in the Taylor expansion of the function $f$ thus bringing us back to TEGR. Equivalently, the $f_{TT}$-term in the equations (\ref{fTeom}) obviously disappears at the linear order.

Given that the full linear theory around the Minkowski background is nothing but simply TEGR, we've got a strong coupling issue in the shape of accidental gauge symmetry. Namely, all the local Lorentz group is fully restored at the linear level. Since it is for sure not the case in general, this is a singular locus of the phase space. Even if, for a moment of desperation, we only cared about propagating modes, there is little doubt that at least one new dynamical degree of freedom must be available in the full model \cite{nontrM}. The rather bad news is that this problematic locus is a very simple and important place for any theory.

In principle, singular loci can be found in phase spaces of many modified gravity theories, like for example at the zeros of the first or the second derivative of the function $f$ in $f(R)$ gravity. What is amazing about $f(\mathbb T)$ theories though, is that the strong coupling issues are really persistent. If not to play with very contrived structures \cite{T02}, the spatially flat cosmology can be built by using a conformally rescaled tetrad $e^a_{\mu}=a(t)\cdot \delta^a_{\mu}$. Long time ago it was noticed that there is still no new dynamical mode in the linear cosmological perturbations around it \cite{cosmong}. An accurate analysis \cite{GKcosm} also shows a bit of accidental gauge symmetry, that for the pseudoscalar mode $s$. In what concerns new dynamical modes, even going for spatially curved cosmologies does not seem to help much \cite{curvcosm}.

\section{Preferred foliations?}

Given this rather unclear situation with the dynamical, constrained and pure gauge modes in the theory, it would be natural to look at the Hamiltonian analysis. It turns out to be a rather complicated endeavour though, due to non-constant ranks of Poisson brackets' algebras of constraints. This is again nothing new for models with ill-defined numbers of degrees of freedom, except for how dense the events of rank changes seem to be in $f(\mathbb T)$ gravity. Actually, there are contradictory results in the literature.

To the best of my knowledge, there are three main Hamiltonian claims \cite{Ham1, Ham2, Ham3} available. In the usual spacetime dimension, the first and the last ones \cite{Ham1, Ham3} found three new dynamical modes, that is on top of the usual two graviton polarisations thus being $2+3=5$ in total, while the middle work \cite{Ham2} insisted on only one new dynamical mode, and therefore three in total. The last work \cite{Ham3} is probably the most accurate one, although there is still no discussion of how the numbers jump in the phase space, and what are the necessary assumptions for getting the full number of claimed modes. Considerations of constant $\mathbb T$ solutions without allowing for variations of that even in perturbations do not count.

The paper \cite{Ham2} which counted less dynamical modes has got an obvious mistake in it. Namely, the spatial derivatives of the auxiliary scalar field equal to $\mathbb T$ had been forgotten in the Poisson brackets \cite{Ham3}. Based on that, it is rather tempting to conclude that the real number of new dynamical modes is three. However, the puzzling point is that, as far as I know, it is not what has ever been seen in explicit perturbative calculations. I've already mentioned just zero new modes in weak gravity and in cosmology. On the other hand, employing the remnant symmetries \cite{remn}, one can have miriads of other soluitons with Minkowski metric and $\mathbb T =0$. By studying perturbations around those, we see at most "almost one" extra mode \cite{nontrM}, with the word "almost" meaning some strange restriction on the freedom of Cauchy data.

How can it be? Recall that the Hamiltonian mistake was in neglecting the spatial gradient of $\mathbb T$. However, unless we go for non-symmetric configurations \cite{T02}, the cosmological spacetimes do mostly have a strictly time-like gradient of it, unless we are talking about bounces or the like. Then one might (partially) fix the gauge of ${\mathbb T}(t,\overrightarrow x)=t$, for both background and perturbations, and then the analysis of that paper \cite{Ham2} seems to apply. If true, it looks like having a preferred foliation in this particular subset of the phase space points. Except for the fact that it is not universal for all possible regimes, it is somewhat similar to the case of cuscuton fields \cite{cusc}.

In order to better see what's going on, let us assume we've fixed a gauge with ${\mathbb T}=t$ and $g_{0i}=0$. In this case the last term in the equations (\ref{fTeom}) takes the form of $f_{TT}\cdot S_{\mu\nu 0}$. It contains at most first derivatives and makes no contribution to the Hamiltonian constraint (the temporal component of the equation). The antisymmetric equations take the form of $S_{\mu\nu 0}-S_{\nu\mu 0}=0$. For the six Lorentzian variables, we get three equations on velocities via ${\mathfrak F}^a_{0i}$ combinations (\ref{comb}) in the mixed components. At the same time, the three other (spatial) antisymmetric equations do only say that $\eta_{ab} e^a_0 {\mathfrak F}^b_{ij}=0$ and ask for no initial data. The three initial data for ${\mathfrak F}^a_{0i}$-equations are also restricted by one condition of ${\mathbb T}=t$, and it all indeed looks like one extra mode in the whole Lorentzian realm.

Note also that it is now more clear what has happened in cosmology \cite{GKcosm}. The scalar part of the  $S_{ij 0}-S_{ji 0}=0$ equation just disappears leading to one equation less, for there is no scalar contribution to $\eta_{ab} e^a_0 {\mathfrak F}^b_{ij}$. There is simply no way one could have an antisymmetric in $i$ and $j$ linear expression for the scalars. For the linear perturbations, the pseudoscalar belongs to an accidental gauge symmetry, for it does not influnce the torsion scalar $\mathbb T$, not even at quadratic level. Intuitively, one can conclude that an extra gauge freedom imposes an extra constraint, and hence no new dynamical modes. Note though that, in the spatially curved cases \cite{curvcosm} at least, there is more to think about.

At the same time, in case of static spherically symmetric solutions, one has a gauge of ${\mathbb T}={\mathbb T}(r)$. Then there is an extra derivative term to the Hamiltonian constraint, and all six antisymmetric equations feature the velocities inside ${\mathfrak F}^a_{0i}$ quantities (\ref{comb}) through $S_{\mu\nu r}-S_{\nu\mu r}$ components. It might very well be about three new dynamical degree of freedom. Unfortunately, it would be hard to explicitly study perturbations around such solutions, for the lack of known exact ones except the rather problematic cases of complex-valued tetrads \cite{compltet}. Probably, a feasible way to go would be to do perturbations around the charged flat-horizon constructions \cite{flattet}, even if much less physical.

Note also that, given such differences for different types of $\mathbb T$ behaviour, there should be no surprise that the Cauchy data might look rather irregular \cite{nontrM} for perturbations around solutions of $\mathbb T=0$, or any other constant value. Going for the constant $\mathbb T$ solutions as for the simplest ones gets to look even more suspicious now.

Finally, I should mention that dynamical issues of $f(\mathbb T)$ were also mentioned already a decade ago \cite{prob1}, in a somewhat nontransparent language of characteristics \cite{prob1}. On top of the "constraint bifurcation", there was also unpredictability of evolution, in the form of an extra gauge freedom. Additional amounts of gauge freedom due to incomplete Lorentz breaking might be good only if one could have them in a stable way and not propagating to the metric sector. Neither the former \cite{obstr} nor the latter \cite{prob2} seem to be the case in modified teleparallel frameworks, at least not in the simplest cases, and therefore the presence of remnant symmetries \cite{remn} does not help us at all \cite{prob3}. In my opinion, we should pay more attention to the papers \cite{prob1, prob2, prob3}, even though I do not buy many of their interpretations.

\section{Discussion and Conclusions}

An interesting lesson to learn is that modifying teleparallel gravity is a very dangerous thing to do. At the same time, very often we make conclusions without a proper ground. One of the most interesting such cases for me is the claims of ghosts in New GR beyond the "one parameter" case \cite{old}. Having put extra derivatives into the action, for the sake of the spin projector formalism, we used to never think about how much it had changed the model at hand.

I think that the paper \cite{them} touches upon a very interesting topic and presents a very important work. It's really time to rethink all the restrictions put on various metric-affine gravity models by blindly relying on the usual quantum-field-theory (QFT) ideas which might not always work. At the same time, the main claim \cite{them} of seeing ghosts in the vector sector doesn't have any proper evidence behind. Basically, every theory with several variables can be rewritten in terms of some constrained modes and another one satisfying a higher derivative equation. One might say that the old paper \cite{old} was more informative in this respect. Even though one cannot imply instability in its way either, it clearly indicated that there would be problems for using the usual QFT techniques in these models.

In my view, the most general New GR is a very promising option because it does have a well-defined number of degrees of freedom. We still have to better study the question of ghosts, and dynamical stability in general. However, more popular models, such as $f(\mathbb T)$, do not have even that. Their strong coupling issues are so ubiquitous that even the number of degrees of freedom is not clear. People still widely use such theories for cosmology, but the problem is very serious. It is of a very doubtful value, to invest much effort in making predictions by closing our eyes at the dynamics being severely ill-defined, to start with.

I'd also like to mention that the discussion of finitely strong and infinitely strong couplings \cite{them} looks rather strange. Even if we make the fine structure constant extremely large, all the equations of electrodynamics do have all the same degrees of freedom: two dynamical, one constrained and one pure gauge. This is a finite case. We only lose our ability of finding solutions perturbatively, let alone going quantum. On the other hand, if we fully lack some modes in a linear analysis, this is an infinitely strong coupling and an ill-defined initial value problem, even though the modes will normally be seen at higher orders \cite{anti}. 

Roughly speaking, at a mostly intuitive level, a coefficient in front of a kinetic term vanishes which means that canonically normalising the field makes a coefficient in a potential term diverge, hence the name. If the coupling is strong but without such singularities, then the equations can be studied mathematically, even if we have no idea of how to make it quantum. Making good sense out of quantum physics is a separate big problem. The issue of $f(\mathbb T)$ is that, even at the purely classical level, it is extremely ill-defined, and it doesn't seem possible to give it some well-defined effective meaning, so that reliable calculations would be somehow available.

\end{document}